\documentclass[12pt,a4paper,british]{iopart}

\usepackage[T1]{fontenc}
\usepackage[latin9]{inputenc}
\usepackage{babel}
\usepackage{graphicx}
\usepackage{subfigure}
\usepackage{epstopdf}
\usepackage[numbers,square,sort&compress]{natbib}
\usepackage[unicode=true,pdfusetitle,bookmarks=false,breaklinks=false,pdfborder={0 0 1},backref=false,colorlinks=false]{hyperref}
\usepackage{array}
\usepackage{verbatim}
\usepackage{amstext}
\usepackage{amssymb}
\usepackage{iopams}
\usepackage{setstack}
\usepackage{soul}

\makeatletter
\@ifundefined{definecolor}{\usepackage{color}}{}

\def\openone{\leavevmode\hbox{\small1\kern-3.8pt\normalsize1}}

\newcommand{\ket}[2][]{{|#2\rangle_{#1}}}
\newcommand{\bra}[2][]{{}_{#1}\langle #2|}

\newcommand{\proj}[2][]{\ket{#2}_{#1}\bra{#2}}
\makeatother

\begin{document}

\title{Dephasing in coherent communication with weak signal states}

\author{Marcin Jarzyna, Konrad Banaszek, Rafa{\l} Demkowicz-Dobrza\'{n}ski}

\address{Instytut Fizyki Teoretycznej, Wydzia{\l} Fizyki, Uniwersytet Warszawski, Ho\.{z}a 69, PL-00-681 Warszawa, Poland}

\ead{Marcin.Jarzyna@fuw.edu.pl, Konrad.Banaszek@fuw.edu.pl, demko@fuw.edu.pl}

\begin{abstract}
We analyze the ultimate quantum limit on the accessible information for an optical communication scheme when time bins carry coherent light pulses prepared in one of several orthogonal modes and the phase undergoes diffusion after each channel use. This scheme, an example of a quantum memory channel, can be viewed as noisy pulse position modulation (PPM) keying with phase fluctuations occurring between consecutive PPM symbols. We derive a general expression for the output states in the Fock basis and implement a numerical procedure to calculate the Holevo quantity. Using asymptotic properties of Toeplitz matrices, we also present an analytic expression for the Holevo quantity valid for very weak signals and sufficiently strong dephasing when the dominant contribution comes from the single-photon sector in the Hilbert space of signal states. Based on numerical results we conjecture an inequality for contributions to the Holevo quantity from multiphoton sectors which implies that in the asymptotic limit of weak signals, for arbitrarily small dephasing the accessible information scales linearly with the average number of photons contained in the pulse. Such behaviour presents a qualitative departure from the fully coherent case.
\end{abstract}



\maketitle

\section{Introduction}

An optical light pulse with a well defined amplitude is represented in the quantum theory by a coherent state prepared in a certain localized mode of the electromagnetic field \cite{glauber}. This imposes restrictions on the possibility of identifying the pulse amplitude, as coherent states are in general not orthogonal in the quantum mechanical sense and consequently cannot be distinguished with certainty \cite{helstrom,wittmann,Tsujino,mueller,Migdall}. The above observation suggests that the capacity of an optical communication link is fundamentally limited by the average optical power and the number of available field modes, which is indeed confirmed by a thorough analysis \cite{caves}. 

The most elementary technique for reading out optical signals is direct detection based on the photoelectric effect, in which the incoming radiation generates a discrete number of counts \cite{Mandel, Rosenberg}. This process does not depend on the phase of the incident light. In many communication schemes direct detection is not optimal, and the readout can be enhanced by coherent phase-sensitive techniques, which combine the incoming signal with a strong coherent reference field and detect resulting superpositions \cite{Cook,Chen}. Such a strategy can approach the quantum limit for certain encodings, but it requires phase stability between the signal and the reference \cite{Ip,Olivares}.

A coherent reference field is therefore an additional resource that allows one to boost the performance of optical communication.
However, coherent detection becomes sensitive to imperfections that direct detection is robust against, such as the phase noise: one can envisage a realistic scenario in which the phase wanders randomly between consecutive transmissions, introducing a mismatch with the reference field.
The purpose of this paper is to analyze how phase noise affects the ultimate quantum limit on the accessible information for a class of optical communication schemes.
Specifically, we consider a sequence of time bins, each of them containing a coherent pulse with a fixed amplitude prepared in one of $M$ distinguishable modes, corresponding to $M$ symbols used for communication. These modes can be thought of as two orthogonal polarizations for $M=2$, or modes with well defined orbital angular momentum in a more general case \cite{molinateriza, leach, angular1, Wang}. In order to explore the intermediate regime between fully coherent detection and the completely dephased limit corresponding to direct detection,
we assume that the phase between consecutive uses of the channel undergoes diffusion described by a canonical model obtained by solving the diffusion equation on a unit circle. General formulas for the transformation of the transmitted states are derived in the multimode Fock basis. The accessible information is estimated with the help of the Holevo bound \cite{holevo1, Holevo2, Westmoreland}. For a finite number of time bins we study analytically the weak amplitude limit when the dominant contribution comes from the one-photon sector. Further, we carry out numerical calculations of the Holevo quantity which allows us to extend the discussion to higher average photon numbers. These results lead us to a conjecture which implies a linear form of the asymptotic scaling in the weak signal limit of the accessible information with the average photon number used for communication. This is a qualitatively different behaviour compared to the fully coherent scheme, when the Holevo quantity as a function of the average photon number becomes infinitely steep when the pulse amplitude tends to zero.

By way of illustration, let us briefly discuss the two extreme cases using the simplest version of the communication scheme that will be studied here. Suppose that two equiprobable symbols, corresponding to two bit values, are encoded as coherent states of equal amplitudes prepared in a single time bin, but in two orthogonal polarizations. If the two polarizations are separated at the output using a polarizing beam splitter and measured using direct detection, the bit value is missing whenever neither of the detectors produces any counts. Such a communication scheme is an example of the well known binary erasure channel \cite{cover}. The probability that one of the detectors clicks, i.e.\ an erasure does not occur, is $1-e^{-\bar{n}}$, where $\bar{n}$ is the average number of photons in the coherent state. This figure gives directly the accessible information for the communication scheme, which for very weak signals, when $\bar{n} \ll 1$, can be approximated as $1-e^{-\bar{n}} \approx \bar{n}$, scaling linearly with $\bar{n}$.

\begin{figure*}[t]\label{fig:dif}
\includegraphics[width=0.75\textwidth]{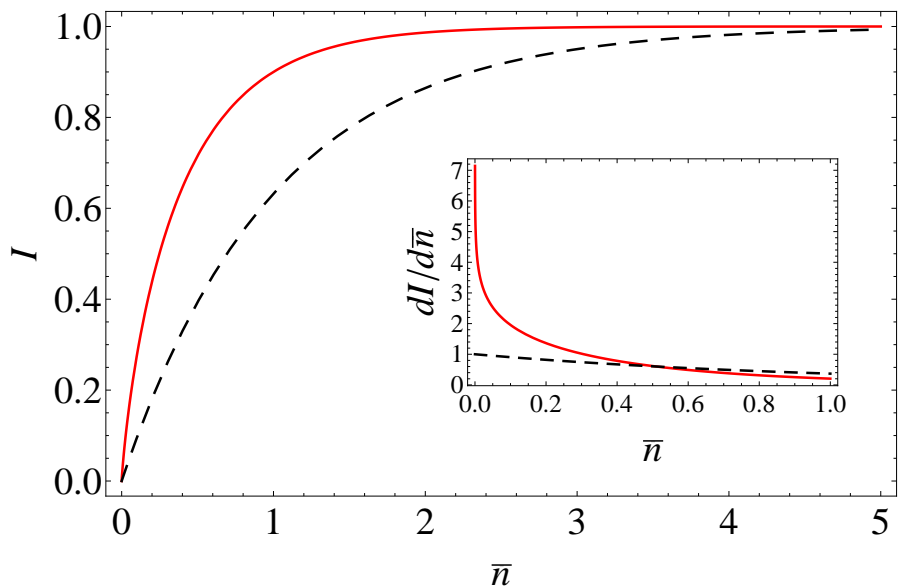}
\caption{Accessible information ${\cal I}$ for encoding information in two orthogonally polarized coherent states with an average photon number $\bar{n}$, assuming incoherent direct detection scheme (dashed black) and the optimal coherent detection saturating the Holevo bound (solid red). The qualitatively different behaviour for $\bar{n} \rightarrow 0$ is illustrated in the inset with a plot of the derivative $d{\cal I}/d\bar{n}$ for both cases.}
\end{figure*}

In contrast, the ultimate quantum limit for classical communication using two equiprobable quantum states depends only on the absolute value of their scalar product, which in our example is $e^{-\bar{n}}$. The accessible information is bounded from above by the Holevo quantity \cite{holevo1, Holevo2, Westmoreland}, given in this specific case by ${\sf H}\bigl( \frac{1}{2} (1-e^{-\bar{n}}) \bigr)$, where ${\sf H} (\cdot)$ stands for binary entropy. This value can be viewed as the maximum accessible information of the communication scheme achievable by implementing collective detection and exploiting phase coherence between consecutive transmissions. For $\bar{n} \ll 1$ the accessible information scales as ${\sf H}\bigl( \frac{1}{2} (1-e^{-\bar{n}}) \bigr) \approx
 \frac{\bar{n}}{2} \log_2 \frac{1}{\bar{n}}$ in the leading order. Consequently, as illustrated in figure \ref{fig:dif} the accesible information becomes infinitely steep when $\bar{n}$ approaches zero, demonstrating the dramatic difference between coherent and incoherent detection. Results presented in this work indicate that for an arbitrarily small amount of phase diffusion this feature is lost and the accessible information scales linearly with $\bar{n}$, albeit it becomes enhanced by a multiplicative factor that diverges to infinity when diffusion becomes negligible. Although this conclusion relies on a numerically motivated conjecture, we have been able to derive an analytical expression for the multiplicative enhancement factor using the Szeg\H{o} limit theorem \cite{szego1, Gray2006}.

The motivation for the model studied in this paper is two-fold. On the practical side, the idea of sending a light pulse in one of $M$ distinguishable modes underlies the pulse position modulation (PPM) format, in which the modes form a temporal train. For very low average photon numbers, the PPM encoding approaches the optimal capacity of a bosonic channel and is currently studied in the context of deep-space communication \cite{Chen,Guha2011,Waseda2011}. Our model can be viewed as a study of dephasing in PPM keying when the phase diffusion occurs only between the consecutive $M$-ary PPM symbols. This assumption is justified when the temporal separation between the symbols is much larger then the overall duration of individual symbols. From the formal perspective, the property of the fixed average photon number per symbol simplifies the mathematical treatment of the model: for a random overall phase, one can analyze separately contributions to the Holevo quantity from individual subspaces with a well-defined total photon number.

This paper is organized as follows. First, in Sec.~\ref{Sec:CommunicationScheme}, we present in details the communication scheme to be studied and set the notation. Sec.~\ref{Sec:Dephasing} introduces the dephasing model. Results for one-photon signals are derived in Sec.~\ref{Sec:One-photon}, and their asymptotics is investigated in Sec.~\ref{Sec:Asymptotic}. In Sec.~\ref{Sec:Numerical} we present approach used in numerical calculations. The numerical results, together with the conjecture on the properties of the Holevo quantity and its implications are discussed in Sec.~\ref{Sec:NResults}. Sec.~\ref{Sec:CollectiveNoise} highlights certain mathematical subtleties of the presented conjecture. Finally, Sec.~\ref{Sec:Conclusions} concludes the paper.

\section{Communication scheme}
\label{Sec:CommunicationScheme}

Transmission of information encoded in a sequence of light pulses with the phase subjected to diffusion can be formally represented as a quantum memory channel \cite{Kretschmann,Mancini}. In this picture, the phase plays the role of a classical memory system that undergoes fluctuations between channel uses and is not affected by channel inputs. We will be interested in a scenario when the initial phase preparation is completely random and neither the initial nor the final value of the phase is available to the communicating parties. In order to analyse the accessible information we will consider communication using blocks of input systems of length $L$. The ultimate quantum bound will be given by the Holevo quantity in the limit of the block size going to infinity.

Specifically, the memory effects will be included by taking as the physical system a block of $L$ time bins, each comprising $M$ orthogonal modes of the electromagnetic radiation, distinguishable for example by their polarization and/or orbital angular momentum \cite{molinateriza, leach, angular1, Wang}. Let $\hat{a}_{lm}$ denote the annihilation operator for the $m$th mode in the $l$th bin, where $l=1,\ldots, L$ and $m=1,\ldots, M$. The orthogonality implies that $[\hat{a}_{lm}, \hat{a}_{l'm'}^\dagger] = \delta_{ll'} \delta_{mm'}$. A single word to be encoded in the block of time bins can be represented as an $L$-element sequence ${\bf m} =(m_1, m_2,\ldots, m_L)$, where each $m_l$, assuming one of $M$ values, specifies the symbol transmitted in the $l$th bin. This word is encoded as the input state
\begin{equation}
\ket{{\bf m}}  = \ket{(\alpha)_{m_1}, (\alpha)_{m_2}, \ldots , (\alpha)_{m_L}}
= e^{-L|\alpha|^2/2} \exp\left( \alpha \sum_{l=1}^{L} \hat{a}^\dagger_{lm_l} \right) \ket{\text{vac}},
\end{equation}
where $\ket{\text{vac}}$ denotes the vacuum state of the electromagnetic field and $\alpha$ is the amplitude of a coherent pulse in an individual time bin, assumed for simplicity to be real.
It is worth noting that $\ket{{\bf m}}$ can be viewed as a coherent state with an amplitude $\sqrt{L}\alpha$ prepared in a delocalized mode characterized by a bosonic operator
\begin{equation}
\hat{b}_{\bf m} = \frac{1}{\sqrt{L}} \sum_{l=1}^{L} \hat{a}_{lm_l}
\end{equation}
which satisfies $[\hat{b}_{\bf m}, \hat{b}_{\bf m}^\dagger] = 1$. We will find this representation helpful when analyzing the Fock state representation of the input states.

It will be convenient to write the state $\ket{{\bf m}}$ as a superposition
\begin{equation}
\ket{{\bf m}} = \sum_{N=0}^{\infty} \sqrt{P_N(L\bar{n})} \ket{\Psi_{\bf m}^{(N)}}.
\end{equation}
Here $\bar{n} = |\alpha|^2$ is the average photon number in an individual time bin,
\begin{equation}
P_N(\mu) = e^{-\mu} \frac{\mu^N}{N!}
\end{equation}
denotes the Poisson distribution with the mean value $\mu$ and $\ket{\Psi_{\bf m}^{(N)}}$ is the normalized $N$-photon component, which can be defined in a compact way as
\begin{equation}
\label{Eq:PsimN}
\ket{\Psi_{\bf m}^{(N)}} = \frac{\bigr(\hat{b}^\dagger_{\bf m}\bigl)^{N}}{\sqrt{N!}} \ket{\text{vac}}.
\end{equation}
The states $\ket{{\bf m}}$ undergo dephasing described by a completely positive trace-preserving map $\Lambda$ whose explicit form will be discussed in detail in Sec.~\ref{Sec:Dephasing}. The strength of the dephasing introduced by $\Lambda$ will be characterized by a real parameter
$\kappa$, to be defined later.
For the time being, the only property required from $\Lambda$ is that it preserves the total photon number $N$ while removing completely coherence between subspaces with different $N$. This allows us to write
\begin{equation}
\label{Eq:Lambda|m>}
\Lambda(\proj{{\bf m}}) = \sum_{N=0}^{\infty} P_N(L\bar{n}) \Lambda (\proj{\Psi_{{\bf m}}^{(N)}}).
\end{equation}

It is assumed that all input states are sent with the same probability $1/M^{L}$. We will be interested in the Holevo quantity
${\cal X}(L,\kappa)$ calculated for a block of $L$ symbols and the dephasing strength $\kappa$, which can be viewed as a bound on the accessible information when the phase is completely randomized after every $L$ uses of the channel.
Explicitly, the Holevo quantity is given by the difference
\begin{equation}
{\cal X}(L,\kappa) = {\sf S}(\Lambda(\hat{\varrho}_{\text{av}})) - \frac{1}{M^L} \sum_{\bf m} {\sf S} (\Lambda(\proj{{\bf m}})),
\end{equation}
between the von Neumann entropy ${\sf S}(\cdot)$ of the averaged state
\begin{equation}
\hat{\varrho}_{\text{av}} = \frac{1}{M^L} \sum_{{\bf m}} \proj{{\bf m}}
\end{equation}
subjected to $\Lambda$ and the average entropy of the individual states $\Lambda(\proj{{\bf m}})$.
Because the states $\Lambda(\proj{\Psi_{{\bf m}}^{(N)}})$ appearing on the right hand side of (\ref{Eq:Lambda|m>}) occupy orthogonal subspaces corresponding to different eigenvalues of the total photon number operator $\hat{N}= \sum_{lm} \hat{a}_{lm}^\dagger \hat{a}_{lm}$, the von Neumann entropy of a state $\Lambda(\proj{{\bf m}})$ can be written as
\begin{equation}
{\sf S}\bigl( \Lambda(\proj{{\bf m}}) \bigr)  = {\sf H} \bigl( \{ P_N(L\bar{n})  \} \bigr) +
\sum_{N=0}^{\infty} P_N(L\bar{n}) {\sf S}\bigl( \Lambda(\proj{\Psi_{{\bf m}}^{(N)}}) \bigr)
\end{equation}
where ${\sf H}\bigl(\{\cdot \} \bigr)$ is the Shannon entropy of a classical probability distribution. Analogously, the von Neumann entropy of the average state after dephasing
$\Lambda(\hat{\varrho}_{\text{av}})$ reads
\begin{equation}
{\sf S}\bigl( \Lambda(\hat{\varrho}_{\text{av}}) \bigr)  = {\sf H} \bigl( \{ P_N(L\bar{n})  \} \bigr) +
\sum_{N=0}^{\infty} P_N(L\bar{n}) {\sf S}\bigl( \Lambda(\hat{\varrho}_{\text{av}}^{(N)}) \bigr)
\end{equation}
where the normalized average density matrix in the $N$-photon sector is given by
\begin{equation}
\hat{\varrho}_{\text{av}}^{(N)} = \frac{1}{M^L} \sum_{{\bf m}} \proj{{\Psi_{{\bf m}}^{(N)}}}.
\end{equation}
Consequently, the Holevo bound for our communication scheme can be written as a weighted sum of contributions from $N$-photon sectors:
\begin{equation}
\label{Eq:Holevo=sum}
{\cal X}(L,\kappa) = \sum_{N=0}^{\infty} P_N(L\bar{n}) {\cal X}^{(N)}(L, \kappa),
\end{equation}
where
\begin{equation}
\label{Eq:XNdef}
{\cal X}^{(N)}(L, \kappa) = {\sf S}\bigr( \Lambda(\hat{\varrho}_{\text{av}}^{(N)})  \bigr)
- \frac{1}{M^L} \sum_{{\bf m}} {\sf S} \bigl( \Lambda(\proj{{\Psi_{{\bf m}}^{(N)}}}) \bigr)
\end{equation}
Let us note that ${\cal X}^{(N)}(L, \kappa)$ are functions of the word length $L$ and depend on the specifics of the dephasing process, but they are independent of the average photon number per time bin $\bar{n}$. This is because each symbol is encoded in a coherent pulse with the same amplitude. The above feature greatly simplifies the analysis of the Holevo quantity.

\section{Dephasing model}
\label{Sec:Dephasing}

Let us now specify details of the dephasing map. We assume that the phases in individual bins are shifted by respective random variables $\phi_1, \ldots \phi_L$ taken from the range $-\pi < \phi_l \le \pi$. The phase $\phi_1$ in the first time bin is completely randomized, while the phase $\phi_l$ for the bin $l=2, \ldots, L$ is given by a conditional probability distribution
\begin{equation}
\label{Eq:CondProbDist}
p(\phi_l | \phi_{l-1}) = p(\phi_l - \phi_{l-1}),
\end{equation}
where
\begin{equation}
\label{Eq:pdiff}
p(\phi) = \frac{1}{2\pi}\left(1+2\sum_{n=1}^\infty\cos(n\phi)e^{-\kappa n^2}\right)
\end{equation}
is a solution of the diffusion equation on the group U(1) with the initial condition in the form of the Dirac delta function $\delta(\phi)$ \cite{Liao2004}. The parameter $\kappa$, playing the role of time in the diffusion equation, defines here the diffusion strength. The characteristic coherence time of our channel, expressed in the number of time bins, is given by $1/\kappa$. For $\kappa \rightarrow 0$ we recover perfectly correlated phases as then $p(\phi)$ approaches the Dirac delta $\delta(\phi)$, while in the limit $\kappa \rightarrow \infty$ we obtain a uniform flat distribution $p(\phi)= \frac{1}{2\pi}$ which corresponds to complete dephasing. The Fourier transform of $p(\phi)$ reads
\begin{equation}
\label{Eq:FTDiffusion}
\int_{-\pi}^{\pi} d\varphi \, p(\varphi) e^{-i m\varphi} = e^{-\kappa m^2}, \qquad m =0, \pm 1, \pm2, \ldots
\end{equation}
This expression will be useful when calculating the transformation of the input states in the Fock basis.

The map $\Lambda$ describing dephasing of the input state for all $L$ time bins can be written explicitly as
\begin{eqnarray}\label{eq:dephmodel}
\fl
\Lambda(\hat{\varrho}) = \frac{1}{2\pi} \int_{-\pi}^{\pi} d\phi_1 \ldots \int_{-\pi}^{\pi}
d\phi_{L} \, p(\phi_2-\phi_1) p(\phi_3 - \phi_2)\ldots
p(\phi_{L}- \phi_{L-1})\nonumber\\
\times  \exp \left( i \sum_{l=1}^{L} \phi_l \hat{n}_l \right) \hat{\varrho}
\exp \left( - i \sum_{l=1}^{L} \phi_l \hat{n}_l \right),
\end{eqnarray}
where $\hat{n}_{l} = \sum_{m=1}^{M} \hat{a}^\dagger_{lm} \hat{a}_{lm}$ is the operator of the total photon number in the $l$th time bin.
Changing integration variables to $\phi_l' = \phi_l - \phi_{l-1}$, $l=2,\ldots, L$, enables one to represent $\Lambda$ as a composition
of commuting maps
$\Lambda = \Lambda_{1} \circ \Lambda_{2} \circ \ldots \circ \Lambda_{L}$, where
\begin{equation}\label{eq:lambda1}
\Lambda_1 (\hat{\varrho}) = \frac{1}{2\pi} \int_{-\pi}^{\pi} d\phi_1 \, \exp \left( i \phi_1\sum_{j=1}^{L}  \hat{n}_{j} \right) \hat{\varrho}
\exp \left( - i \phi_1 \sum_{j=1}^{L}  \hat{n}_{j} \right)
\end{equation}
and
\begin{equation}
\Lambda_l (\hat{\varrho}) = \int_{-\pi}^{\pi} d\phi'_l \, p(\phi'_l) \exp \left( i \phi'_l\sum_{j=l}^{L}  \hat{n}_{j} \right) \hat{\varrho}
\exp \left( - i \phi_l' \sum_{j=l}^{L}  \hat{n}_{j} \right)
\end{equation}
for $l=2, \ldots , L$.

If the input state is represented as a density matrix in the Fock states basis for the set of modes $\hat{a}_{lm}$, then the action of $\Lambda$
results simply in multiplying off-diagonal elements by certain coefficients. The explicit form of these coefficients can be found as follows.
Let $\ket{\Phi}$ and $\ket{\Phi'}$ be two states such that the number of photons in each time bin is fixed:
\begin{equation}
\hat{n}_l \ket{\Phi} = n_l \ket{\Phi} , \quad
\hat{n}_l \ket{\Phi'} = n'_l \ket{\Phi'}, \qquad l = 1,2,\ldots, L.
\end{equation}
It will be convenient to denote sequences of photon numbers in individual time bins as ${\bf n} = (n_1, \ldots n_L)$ and ${\bf n}' = (n_1', \ldots n_L')$.
Obviously, averaging with respect to the overall phase defined in (\ref{eq:lambda1}) implies that $\Lambda_1(\ket{\Phi}\bra{\Phi'})$ is non-zero only if the total photon numbers in both the states are identical,
$\sum_{l=1}^L n_l = \sum_{l=1}^L n'_l$. We will restrict our attention to this case for the remainder of the present section. Then the Fourier transform given in (\ref{Eq:FTDiffusion}) yields:
\begin{equation}
\Lambda_l \bigl( \ket{\Phi}\bra{\Phi'} \bigr) = \exp\left[ -\kappa \left( \sum_{j=l}^{L} n_j - n'_j \right)^2 \right]\ket{\Phi}\bra{\Phi'}.
\end{equation}
Consequently, for the composition of all $L$ maps $\Lambda = \Lambda_{1} \circ \Lambda_{2} \circ \ldots \circ \Lambda_{L}$ we have
\begin{equation}
\Lambda \bigl( \ket{\Phi}\bra{\Phi'} \bigr) = \exp(-\kappa\lambda_{{\bf n}, {\bf n'}}) \ket{\Phi}\bra{\Phi'},
\end{equation}
where the dephasing factor $\lambda_{{\bf n}, {\bf n'}}$ appearing in the exponent can be written as
\begin{equation}
\lambda_{{\bf n}, {\bf n'}} = \sum_{l=2}^{L} \left( \sum_{j=l}^{L} (n_j - n'_j) \right)^2 = - \sum_{l=2}^{L}
\left( \sum_{i=1}^{l-1} (n_i - n'_i) \right)
\left( \sum_{j=l}^{L} (n_j - n'_j) \right).
\end{equation}
In the second equality we applied to one of the two identical sums over $j$ in the squared large round bracket the fact that $\sum_{j=1}^{L} (n_j - n'_j) = 0$.
A simple counting how many times a product $(n_i-n'_i)(n_j - n'_j)$ for a given pair of indices $i$ and $j$ appears in the above sums yields the expression
\begin{equation}
\label{Eq:lambda=sumfinal}
\lambda_{{\bf n}, {\bf n'}} =  - \sum_{i=1}^{L} \sum_{j=i+1}^{L} (j-i) (n_i-n'_i)(n_j - n'_j).
\end{equation}
It is seen that the contributions to the dephasing factor grow linearly with the distance between the time bins. This reflects the fact that a convolution of conditional probability distributions defined in (\ref{Eq:CondProbDist}) has also the form of (\ref{Eq:pdiff}) with the parameter $\kappa$ multiplied by an integer specifying the distance between the time bins.

\section{One-photon sector}
\label{Sec:One-photon}

Obviously, the zero-photon sector contribution ${\cal X}^{(0)}(L, \kappa) $ to the Holevo quantity in (\ref{Eq:Holevo=sum}) is identically equal to zero. Therefore for very weak signals, when $L\bar{n} \ll 1$, i.e.\ the average number of photons in the sequence of $L$ time bins is much less than one, the Poisson distribution $P_N(L\bar{n})$ in (\ref{Eq:Holevo=sum}) assigns the largest weight to the one-photon contribution ${\cal X}^{(1)}$.
Because the contribution to the Holevo quantity from any $N$-photon sector cannot exceed ${\cal X}^{(N)}(L,\kappa) \le L \log_2 M$ which corresponds to perfect decoding of the symbols, we expect that for sufficiently weak signals the Holevo quantity will be well approximated by
\begin{equation}
\label{Eq:Xonephotonapprox}
{\cal X}(L,\kappa) \approx L \bar{n} {\cal X}^{(1)}(L,\kappa).
\end{equation}
In this section we will analyze in detail the structure of input states in the one-photon sector in order to find a closed expression for ${\cal X}^{(1)}(L,\kappa)$.

It will be convenient to introduce in the one-photon sector of the Hilbert space a formal decomposition into the time bin subsystem $\tau$ spanned by $L$ states $\ket[\tau]{l}$ and the symbol subsystem $\sigma$ spanned by $M$ states $\ket[\sigma]{m}$ such that
\begin{equation}
\hat{a}_{lm}^\dagger \ket{\text{vac}} = \ket[\tau]{l} \otimes \ket[\sigma]{m}, \qquad l =1, \ldots , L, \quad m=1,\ldots , M.
\end{equation}
The normalized density matrix for an individual input state in the one-photon sector
can be written as
\begin{equation}
\label{Eq:Psi(1)mDecomp}
\proj{\Psi^{(1)}_{{\bf m}}} = \frac{1}{L} \sum_{l,l'=1}^{L} \ket[\tau]{l}\bra{l'} \otimes \ket[\sigma]{m_l}\bra{m_{l'}},
\end{equation}
which after dephasing is transformed into
\begin{equation}
\Lambda(\proj{\Psi^{(1)}_{{\bf m}}}) = \frac{1}{L} \sum_{l,l'=1}^{L} e^{-\kappa|l-l'|} \ket[\tau]{l}\bra{l'} \otimes \ket[\sigma]{m_l}\bra{m_{l'}}.
\end{equation}
It is straightforward to notice that the entropy of this state is equal to the entropy of a density matrix for the subsystem $\tau$ alone obtained by replacing $\ket[\tau]{l}\bra{l'} \otimes \ket[\sigma]{m_l}\bra{m_{l'}}$ with $\ket[\tau]{l}\bra{l'}$. The latter  can be in turn written as ${\sf S}(\frac{1}{L}\hat{T})$, where
\begin{equation}
\label{Eq:Tdef}
\hat{T} = \sum_{l,l'=1}^{L} e^{-\kappa|l-l'|} \ket[\tau]{l}\bra{l'}
\end{equation}
is an example of a so-called Toeplitz matrix \cite{Gray2006}, for which elements depend only on the difference of the indices $l-l'$.
The von Neumann entropy ${\sf S}\bigl( \Lambda(\proj{\Psi^{(1)}_{{\bf m}}})\bigr)$ has the same value for any word ${\bf m}$.

Let us now inspect the average input density matrix $\hat{\varrho}_{\text{av}}^{(1)}$ in the one-photon sector, which using (\ref{Eq:Psi(1)mDecomp}) can be written as:
\begin{equation}
\hat{\varrho}_{\text{av}}^{(1)} = \frac{1}{L} \sum_{l,l'=1}^{L} \ket[\tau]{l}\bra{l'}
\otimes \left(\frac{1}{M^L} \sum_{{\bf m}}  \ket[\sigma]{m_l}\bra{m_{l'}} \right)
\end{equation}
The average of the rank one operator $\ket[\sigma]{m_l}\bra{m_{l'}}$ over all words ${\bf m}$ depends on whether the indices $l$ and $l'$ are equal or different. If $l=l'$, the average is given by $\frac{1}{M} \hat{\openone}_\sigma$ where
$\hat{\openone}_\sigma= \sum_{m=1}^{M} \proj[\sigma]{m}$
is the identity operator in the subspace $\sigma$, while for $l \neq l'$ the average reads $\frac{1}{M} \proj[\sigma]{s}$,
where $\ket[\sigma]{s} = \frac{1}{\sqrt{M}} \sum_{m=1}^{M} \ket[\sigma]{m}$ is an equally weighted superposition of all symbol states. Therefore, the one-photon sector of the averaged input state can be written as
\begin{equation}
\hat{\varrho}_{\text{av}}^{(1)} = \frac{1}{LM} \sum_{l,l'=1}^{L} \ket[\tau]{l}\bra{l'} \otimes
[ \proj[\sigma]{s} + \delta_{ll'} (\hat{\openone}_\sigma - \proj[\sigma]{s} )]
\end{equation}
As a result, the average state at the output reads
\begin{equation}
\Lambda(\hat{\varrho}_{\text{av}}^{(1)} ) = \frac{1}{LM} \hat{T} \otimes \proj[\sigma]{s}
+ \frac{1}{LM} \hat{\openone}_{\tau} \otimes (\hat{\openone}_\sigma - \proj[\sigma]{s} )
\end{equation}
where $\hat{T}$ has been defined in (\ref{Eq:Tdef}) and $\hat{\openone}_{\tau}= \sum_{l=1}^{L} \proj[\tau]{l}$ is the identity operator for the time bin subsystem $\tau$.
The two terms in the above equation contain orthogonal projectors $\proj[\sigma]{s}$ and $\hat{\openone}_\sigma - \proj[\sigma]{s}$ in the symbol subspace $\sigma$. Consequently, the von Neumann entropy ${\sf S}(\Lambda(\hat{\varrho}_{\text{av}}^{(1)} ))$
can be simplified to the form
\begin{equation}
{\sf S}\bigl( \Lambda(\hat{\varrho}_{\text{av}}^{(1)} ) \bigr) = \log_2 M + \frac{M-1}{M}  \log_2 L + \frac{1}{M}
{\sf S}({\textstyle\frac{1}{L}}\hat{T}).
\end{equation}
This yields
\begin{eqnarray}
{\cal X}^{(1)}(L,\kappa)  & = & {\sf S}\bigl( \Lambda(\hat{\varrho}_{\text{av}}^{(1)} ) \bigr) -
{\sf S} ( {\textstyle\frac{1}{L}}\hat{T}) =
\log_2 M + \frac{M-1}{M}  [ \log_2 L - {\sf S}({\textstyle\frac{1}{L}}\hat{T}) ] \nonumber \\
\label{Eq:X1sumh(t_i)}
&  = & \log_2 M + \frac{M-1}{LM} \sum_{i=1}^{L} h(t_i),
\end{eqnarray}
where $h(x) = x \log_2 x$ and $t_i$ are eigenvalues of the matrix $\hat{T}$. We will analyze this expression, including the asymptotic limit $L\rightarrow \infty$, in the next section.

\section{Asymptotic limit}
\label{Sec:Asymptotic}

Although the assumption $L\bar{n} \ll 1$ made in (\ref{Eq:Xonephotonapprox}) requires $L$ to be finite, let us inspect the asymptotic limit of (\ref{Eq:X1sumh(t_i)}) when $L\rightarrow \infty$. We will find a closed analytic expression in this limit and show that it approximates with good accuracy the value of ${\cal X}^{(1)}(L,\kappa)$ even for finite word lenghts $L$ provided that dephasing is sufficiently strong.

\begin{figure*}
\centering
\subfigure[]{
\label{fig:chi1}\includegraphics[angle=0,width=0.47\textwidth]{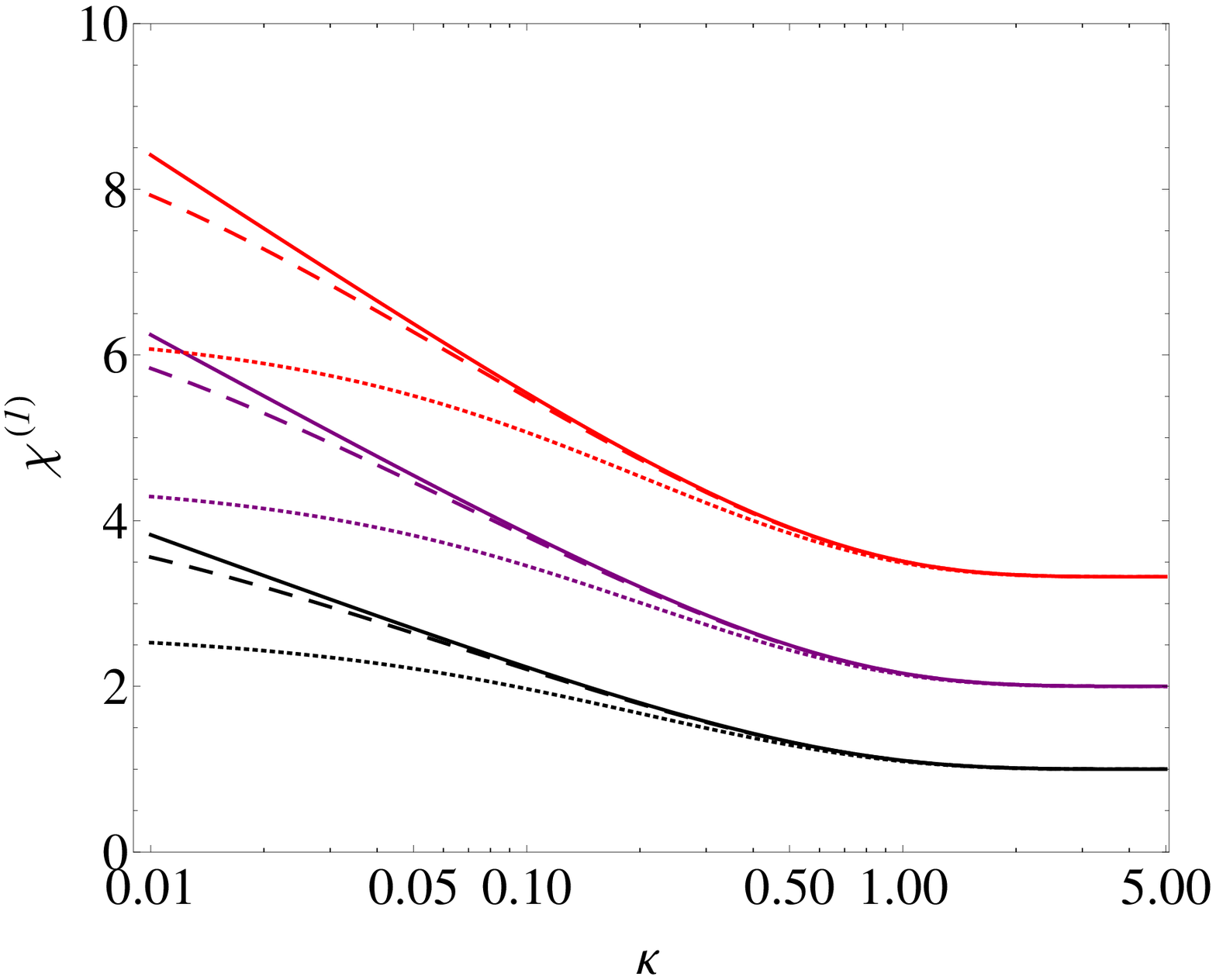}}
\subfigure[]{
\label{fig:chi1contour}\includegraphics[width=0.47\textwidth]{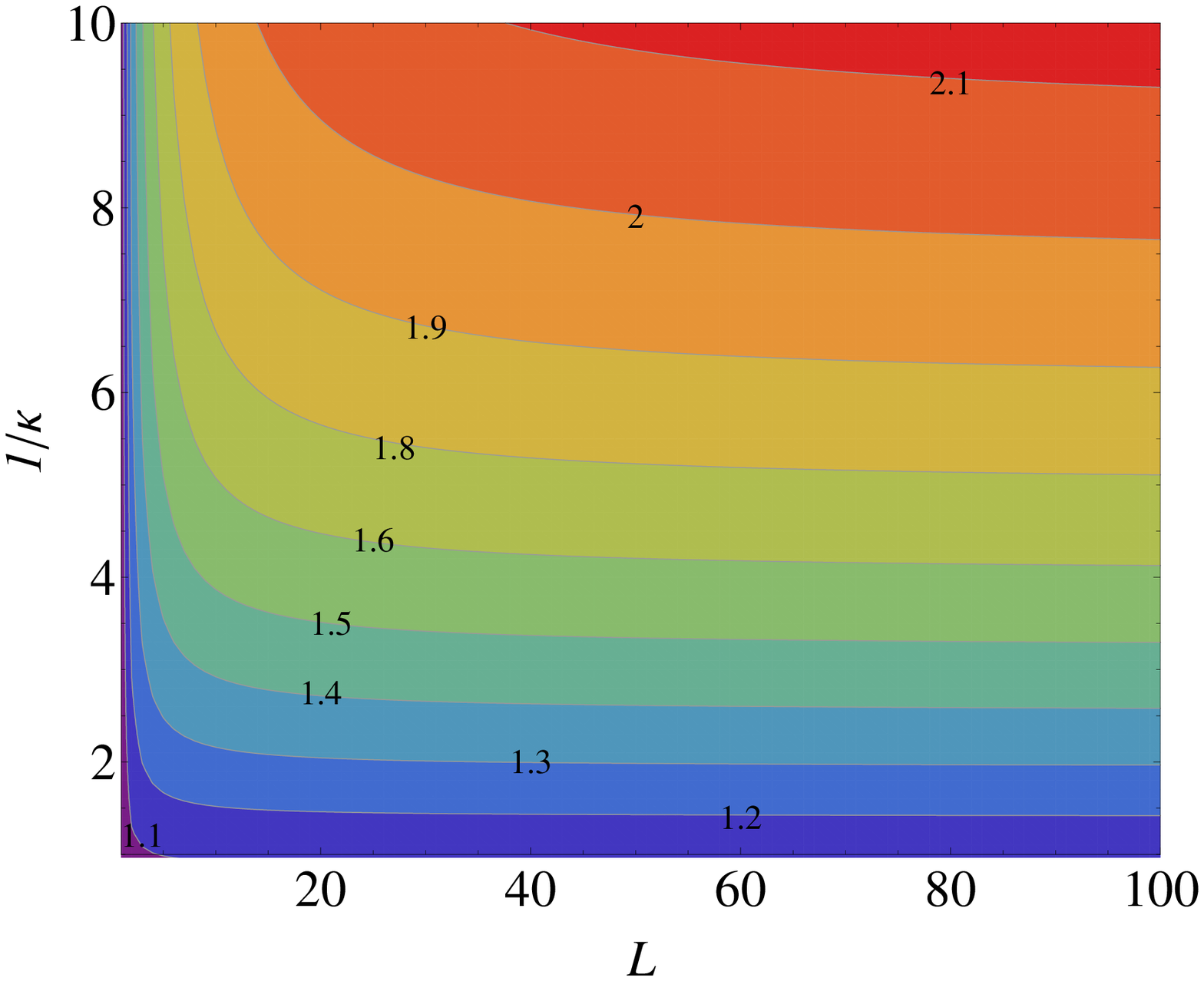}}
\caption{(a) The asymptotic value ${\cal X}^{(1)}_{L\rightarrow \infty}(\kappa)$ as a function of $\kappa$ for codings with $M=2$ (solid black), $M=4$ (solid purple), and $M=10$ (solid red) symbols. One-photon contributions for finite time bin sequences ${\cal X}^{(1)}(L,\kappa)$ are shown for $L=10$ (dotted) and $L=100$ (dashed) using the same colours. (b)
Contour plot of ${\cal X}^{(1)}(L,\kappa)$ as a function of the codeword length $L$ and the inverse of the dephasing constant $1/\kappa$ for binary coding with $M=2$.}
\end{figure*}

The crucial feature of the derived expression for ${\cal X}^{(1)}(L,\kappa)$ that enables further simplification is that $t_i$ are eigenvalues of a Toeplitz matrix whose elements depend only on the distance from the diagonal, $\bra{i+n}\hat{T}\ket{i}=T_{n}$. For this class of matrices we can use the Szeg\H{o} theorem \cite{szego1, Gray2006}, which states that if the series $\sum_{n=0}^{L-1} |T_n|$ remains finite with the increasing matrix size when $L \rightarrow \infty$, then in this limit the average of the values
 of any continuous function $h(\cdot)$ on the eigenvalues of the matrix $\hat{T}$ approaches an asymptotic value given by an integral
\begin{equation}\label{eq:integral}
\frac{1}{L} \sum_{i=1}^{L} h(t_i)  \longrightarrow \frac{1}{2\pi} \int_{-\pi}^{\pi} d\theta \, h\bigl(f(\theta)\bigr),
\end{equation}
where
\begin{equation}
f(\theta) = \sum_{n=-\infty}^{\infty} T_n e^{i n \theta}
\end{equation}
is the inverse Fourier transform of the coefficients $T_n$. Szeg\H{o} theorem holds for any Toeplitz matrix as long as the sum of absolute values $\sum_{n=-\infty}^{\infty} |T_n| $ is finite.
In our case $T_n = e^{-\kappa |n|}$ and the above series is convergent for any $\kappa > 0$. It is also straightforward to calculate a closed expression for
$f(\theta)$:
\begin{equation}
\label{Eq:f(theta)}
f(\theta) = \frac{1-e^{-2\kappa}}{1+e^{-2\kappa} - 2 e^{-\kappa} \cos\theta}.
\end{equation}
Further, we show in \ref{Sec:AppA} that the integral on the right hand side of (\ref{eq:integral}) with the function $f(\theta)$ given by (\ref{Eq:f(theta)}) can be evaluated analytically. This yields finally the following asymptotic expression for the Holevo quantity in the one-photon sector:
\begin{equation}\label{eq:chi1}
{\cal X}^{(1)}_{L\rightarrow \infty}(\kappa) = \log_2 M - \frac{M-1}{M}\log_2\left(1-e^{-2\kappa}\right).
\end{equation}
In figure \ref{fig:chi1} we depict ${\cal X}^{(1)}_{L\rightarrow \infty}$ given by  (\ref{eq:chi1}) as a function of $\kappa$ for several numbers $M$ of symbols used for communication. When $\kappa \rightarrow \infty$, i.e.\ coherence between consecutive bins is erased, the proportionality factor approaches the value $\log_2 M$. This factor, multiplied by the average photon number $\bar{n}$ gives the standard expression for the capacity of an $M$-ary erasure channel in the regime of high erasure probability. For finite $\kappa$ the proportionality factor increases, as the coherence that survives dephasing allows one to boost the channel capacity. In the same graph, we also show ${\cal X}^{(1)}(L,\kappa)$ calculated for finite numbers of time bins, $L=10$ and $L=100$. It is seen that ${\cal X}^{(1)}(L,\kappa)$ deviates from ${\cal X}^{(1)}_{L\rightarrow \infty}(\kappa)$ towards lower values for small $\kappa$, as restricting the length of the time bin sequence prevents one from exploiting fully the state coherence at the detection stage.


In order to gain insight into the validity of the limiting expression ${\cal X}^{(1)}_{L\rightarrow \infty}$ obtained from Szeg\H{o} theorem, in figure \ref{fig:chi1contour} we present a contour plot of
${\cal X}^{(1)}(L,\kappa)$ calculated according to (\ref{Eq:X1sumh(t_i)}) as a function of the number of bins $L$ and the channel coherence time $1/\kappa$. The graph indicates that the limiting value offers a good approximation in the region when $L \gg 1/\kappa$. Let us recall that truncating the Holevo quantity to the one-photon sector is valid when $L \ll 1/\bar{n}$. Combining these two inequalities gives a condition $\bar{n}/\kappa \ll 1$, i.e.\ the average number of photons transmitted over the coherence time of the channel needs to be much smaller than one for the limiting value ${\cal X}^{(1)}_{L\rightarrow \infty}(\kappa)$  to be reached.

Finally, let us point out that the expression given in (\ref{eq:chi1}) diverges as $\kappa \rightarrow 0$. This is easily understood, as for finite $L$ and $\kappa =0$
the matrix $\hat{T}$ has only one non-zero eigenvalue equal to $L$ which gives
\begin{equation}\label{eq:kappa0}
{\cal X}^{(1)}_{\kappa=0}=\log_2 M +\frac{M-1}{M}\log_2 L.
\end{equation}
At the same time, the off-diagonal elements of the Toeplitz matrix $\hat{T}$ are all equal to one, $T_n =1$, and the assumptions of the Szeg\H{o} theorem are not satisfied.


\section{Numerical approach}
\label{Sec:Numerical}

In order to go beyond the one-photon approximation and find contributions to the Holevo quantity defined in (\ref{Eq:XNdef}) from sectors containing $N>1$ photons we need to resort to numerical calculations. We used the following computational approach to obtain numerical results.

The first observation is that for a fixed photon number $N$ the entropies of individual states ${\sf S} \bigl( \Lambda(\proj{{\Psi_{{\bf m}}^{(N)}}}) \bigr)$ are all equal for any word ${\bf m}$. It is therefore sufficient to consider a single word ${\bf m}$ in order to calculate the second term in (\ref{Eq:XNdef}). Starting from the definition given in (\ref{Eq:PsimN}) and using multinomial expansion of the $N$th power $(\hat{b}_{\bf m}^\dagger)^{N}$ of the respective creation operator the input state $\ket{\Psi^{(N)}_{\bf m}}$ can be written as
\begin{equation}
\ket{\Psi^{(N)}_{\bf m}} = \sqrt{\frac{N!}{L^N}} \sum_{{\bf n}}
\frac{(\hat{a}_{1m_1}^\dagger)^{n_1} \ldots (\hat{a}_{Lm_L}^\dagger)^{n_L}}{n_1! \ldots n_L!} \ket{\text{vac}}
\end{equation}
Here the summation is carried out over all sequences of nonnegative integers ${\bf n} = (n_1, \ldots , n_L)$ satisfying the constraint
$\sum_{l=1}^{L} n_l = N$. It will be helpful to denote multimode Fock states present in the superposition as
\begin{equation}
\label{Eq:ketnmdef}
\ket{{\bf n}; {\bf m}} = \frac{(\hat{a}_{1m_1}^\dagger)^{n_1} \ldots (\hat{a}_{Lm_L}^\dagger)^{n_L}}{\sqrt{n_1! \ldots n_L!}} \ket{\text{vac}}.
\end{equation}
The density matrix of the output state written in the Fock basis
\begin{equation}
\Lambda(\proj{{\Psi_{{\bf m}}^{(N)}}}) = \sum_{{\bf n}, {\bf n}'} q_{{\bf n}, {\bf n}'} \ket{{\bf n}; {\bf m}} \bra{{\bf n}'; {\bf m}}
\end{equation}
has elements given by
\begin{equation}
q_{{\bf n}, {\bf n}'} = \frac{N!}{L^N} \frac{\exp(-\kappa \lambda_{{\bf n}, {\bf n}'})}{\sqrt{{n_1! \ldots n_L! {n_1'! \ldots n_L'!}}}}
\end{equation}
where $\lambda_{{\bf n}, {\bf n}'}$ are given by (\ref{Eq:lambda=sumfinal}). In order to estimate the computational complexity of the eigenproblem that needs to be solved to calculate the von Neumann entropy ${\sf S}(\Lambda(\proj{{\Psi_{{\bf m}}^{(N)}}}))$, let us find the number of vectors $\ket{{\bf n}; {\bf m}}$ for a given total number of photons $N$ and a fixed word ${\bf m}$. For this purpose we need to calculate in how many distinguishable ways $N$ photons can be distributed among $L$ time bins. A number $N$ can be written as a composition (i.e.\ an ordered partition) of $k$ greater than zero integers in
${N-1 \choose k-1}$ ways. Time bins for these integers can be chosen in ${L \choose k}$ ways. This gives altogether
$\sum_{k=1}^{\min(L,N)} {N-1 \choose k-1} {L \choose k} $ distinguishable states.

Let us now consider the density matrix of the average output state. It is given explicitly by
\begin{equation}
\label{Eq:varrhoNav=nm}
\Lambda(\hat{\varrho}^{(N)}_{\text{av}}) = \frac{1}{M^L} \sum_{{\bf n}, {\bf n}'} q_{{\bf n}, {\bf n}'} \sum_{{\bf m}} \ket{{\bf n}; {\bf m}} \bra{{\bf n}'; {\bf m}}.
\end{equation}
Here the first  sum is over sequences ${\bf n}$ and ${\bf n}'$ satisfying the constraint $\sum_{l=1}^L n_l = \sum_{l=1}^L n_l' =N$,
while the second sum is carried out over all $M^L$ words ${\bf m}$. Here attention needs to be paid to the number of distinguishable states $\ket{{\bf n}; {\bf m}}$ for a given sequence ${\bf n}$. Of course, if two words ${\bf m}$ and ${\bf m}'$ differ only for time bins in which no photons are present, i.e.\ $m_l \neq {m'}_l$ occurs only for $l$ such that $n_l=0$, then the kets $\ket{{\bf n}; {\bf m}}$ and $\ket{{\bf n}; {\bf m}'}$ correspond to the same physical state according to the definition in (\ref{Eq:ketnmdef}). Therefore for a fixed ${\bf n}$ we will have only $M^{k({\bf n})}$ distinguishable states, where $k({\bf n})$ is the number of non-zero entries in the sequence ${\bf n}$. We will label these distinguishable states as $\ket{{\bf n}; {\boldsymbol \mu}}$, where ${\boldsymbol \mu}$ is a sequence of length $k({\bf n})$ specifying symbols only for time bins occupied by at least one photon.

In order to avoid the superfluous degeneracy associated with states $\ket{{\bf n}; {\bf m}}$ we will write the average output state $\Lambda(\hat{\varrho}^{(N)}_{\text{av}})$ in the basis states $\ket{{\bf n}; {\boldsymbol \mu}}$:
\begin{equation}
\Lambda(\hat{\varrho}^{(N)}_{\text{av}}) = \sum_{{\bf n}, {\bf n}'} \sum_{{\boldsymbol\mu}, {\boldsymbol\mu}'}
\bar{q}_{{\bf n}; {\boldsymbol\mu}, {\bf n}'; {\boldsymbol\mu}'} \ket{{\bf n}; {\boldsymbol \mu}} \bra{{\bf n}'; {\boldsymbol \mu}'}
\end{equation}
and give formulas for the coefficients $\bar{q}_{{\bf n}; {\boldsymbol\mu}, {\bf n}'; {\boldsymbol\mu}'}$.
If ${\bf n} = {\bf n}'$, (\ref{Eq:varrhoNav=nm}) implies that only elements with ${\boldsymbol\mu} = {\boldsymbol\mu}'$ are nonzero. A given ${\boldsymbol\mu}$ is realized by $M^{L-k({\bf n})}$ terms in the sum over ${\bf m}$ in (\ref{Eq:varrhoNav=nm}). Consequently,
\begin{equation}
\bar{q}_{{\bf n}; {\boldsymbol\mu}, {\bf n}; {\boldsymbol\mu}'} = \frac{q_{{\bf n}, {\bf n}}}{M^L} M^{L-k({\bf n})}\delta_{{\boldsymbol\mu}, {\boldsymbol\mu}'}
= \frac{N! \delta_{{\boldsymbol\mu}, {\boldsymbol\mu}'}}{L^N M^{k({\bf n})}} \prod_{l=1}^{L} \frac{1}{n_l!}
\end{equation}
Let us now consider the density matrix elements when ${\bf n} \neq {\bf n}'$. If the symbols specified by ${\boldsymbol\mu}$ and ${\boldsymbol\mu}'$ differ at (at least one) time bin for which both states $\ket{{\bf n}; {\boldsymbol \mu}}$ and $\ket{{\bf n}'; {\boldsymbol \mu}'}$ contain nonzero photons,
then $\bar{q}_{{\bf n}; {\boldsymbol\mu}, {\bf n}'; {\boldsymbol\mu}'} = 0$ as the corresponding term of the form $\ket{{\bf n}; {\boldsymbol \mu}}\bra{{\bf n}'; {\boldsymbol \mu}'}$ never appears in the sum over ${\bf m}$ given in (\ref{Eq:varrhoNav=nm}). If the symbols are identical at all the time bins that contain non-zero photons for both ${\bf n}$ and ${\bf n}'$, we need to count how many times this instance would occur in the summation over ${\bf m}$ in (\ref{Eq:varrhoNav=nm}). Because symbols are fixed in time bins for which either ${\bf n}$ or ${\bf n}'$ give a non-zero photon number, we are free only to choose symbols in time bins in which both ${\bf n}$ and ${\bf n}'$ are zero. The number of these time bins is equal to $L- k({\bf n} + {\bf n}')$. Consequently, the corresponding density matrix element is given by
\begin{equation}
\bar{q}_{{\bf n}; {\boldsymbol\mu}, {\bf n}'; {\boldsymbol\mu}'} = \frac{q_{{\bf n}, {\bf n}'}}{M^{L}}M^{L-k({\bf n}+ {\bf n}')}
=
\frac{N! \exp(-\kappa \lambda_{{\bf n}, {\bf n}'})}{M^{k({\bf n}+ {\bf n}')}} \prod_{l=1}^{L} \frac{1}{\sqrt{n_l!n_l'!}}.
\end{equation}

The number of distinguishable states $\ket{{\bf n}; {\boldsymbol \mu}}$ spanning the average output state can be calculated similarly as before for an individual output state, but in the current case we need to take into account all possible combinations of symbols for a given distribution of photons between time bins.
When $N$ is represented as a composition of $k$ integers greater than zero, we will now have $M^k$ states distinguishable by chosen symbols. Consequently, the total number of
states that need to be taken into account is $\sum_{k=1}^{\min(L,N)} {N-1 \choose k-1} {L \choose k} M^k$. This effectively limits the number of time bins and photons that can be taken into account in numerical calculations. As an example, for binary encoding $M=2$ and $L=4$ time bins the dimension of the relevant Hilbert space is 952 for $N=7$ and 1408 for $N=8$.

\section{Numerical results}
\label{Sec:NResults}

\begin{figure*}
\centering
\subfigure[]{
\label{fig:chi1L4}\includegraphics[angle=0,width=0.47\textwidth]{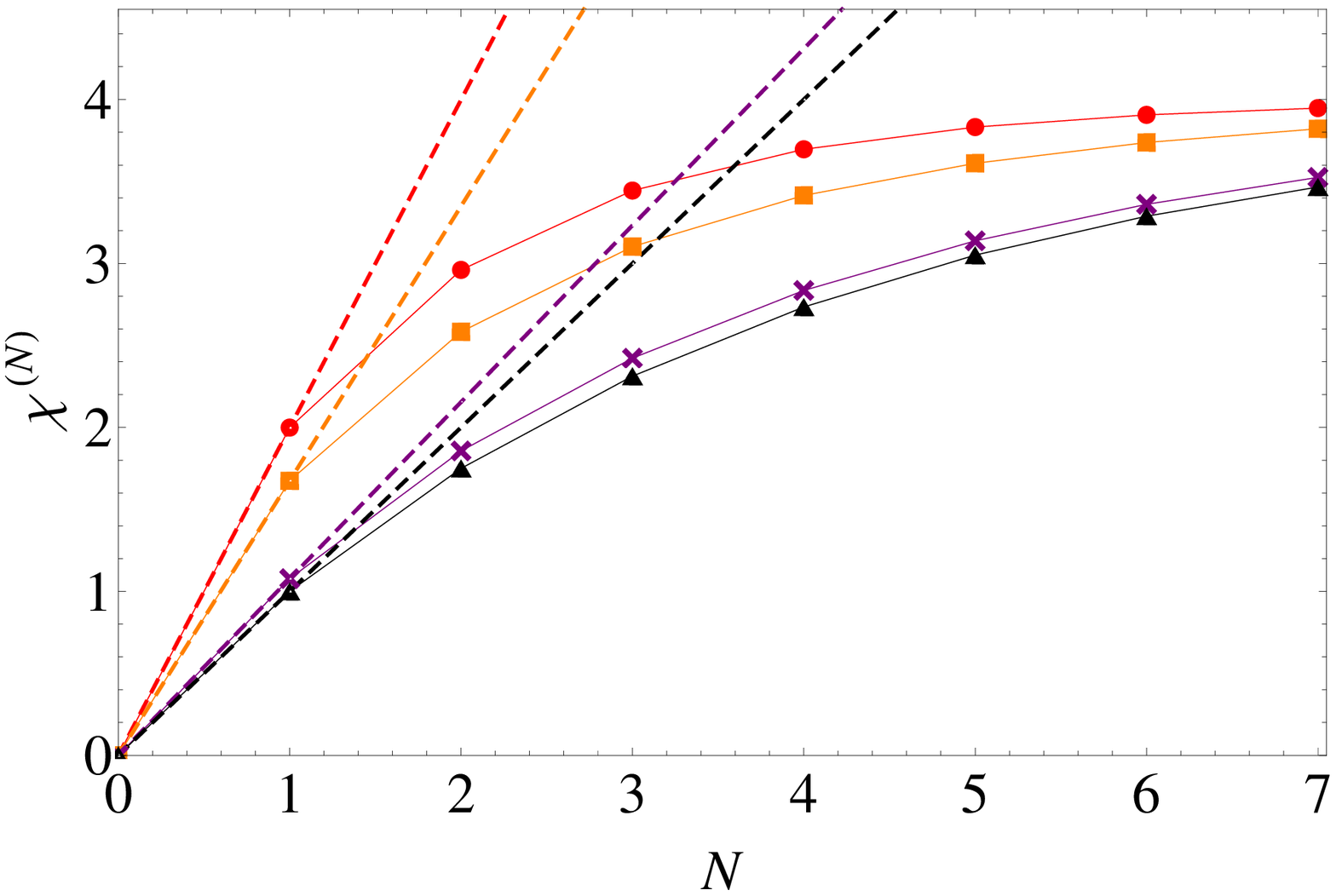}}
\subfigure[]{
\label{fig:chi1L5}\includegraphics[width=0.47\textwidth]{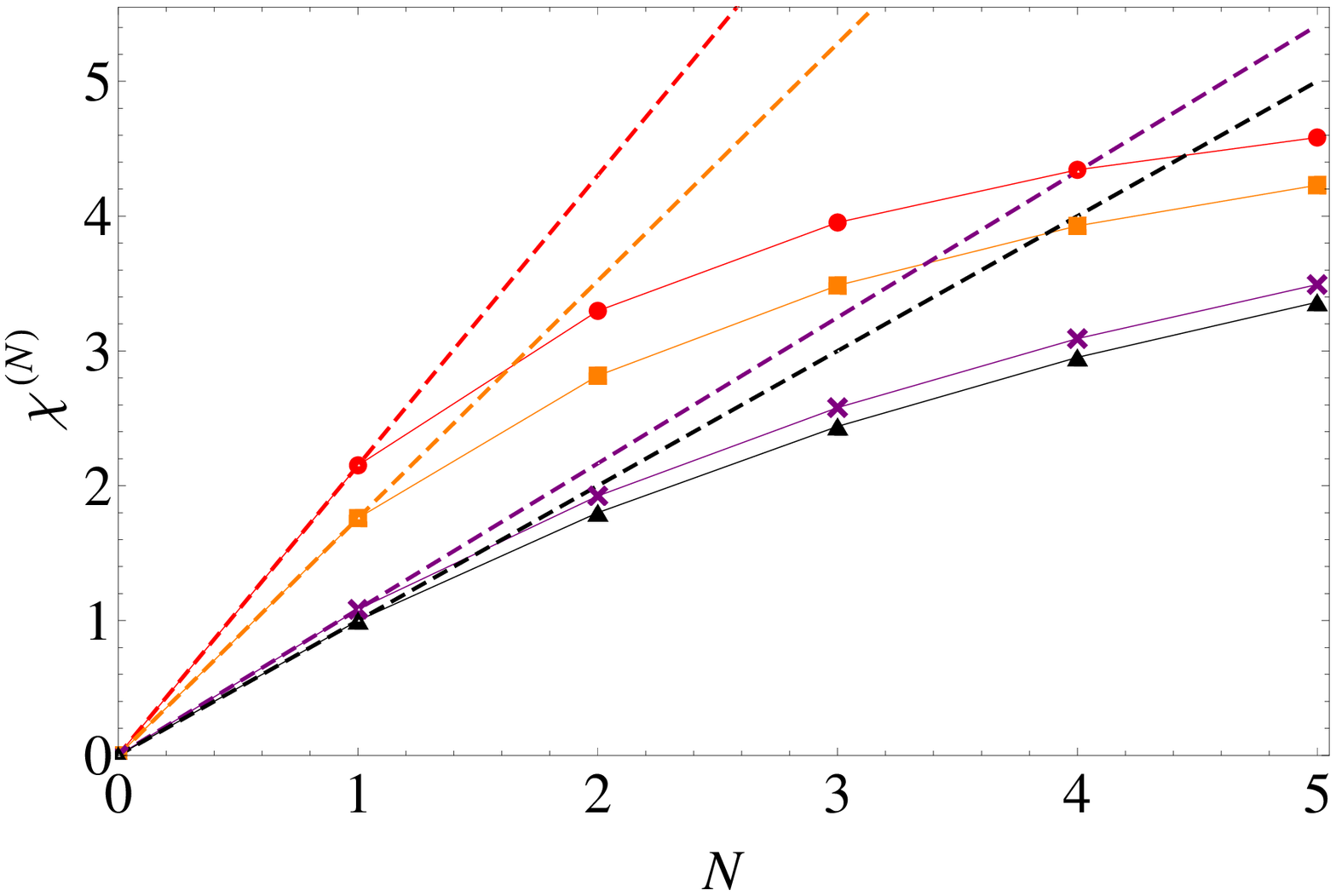}}
\caption{\label{fig:chiN}
Contributions ${\cal X}^{(N)}(L,\kappa)$ to the Holevo quantity from $N$-photon sectors assuming binary encoding $M=2$ and (a) $L=4$ and (b) $L=5$ time bins for the dephasing parameter $\kappa=0$ (circles, red), $\kappa=0.1$ (squares, orange), $\kappa=1$ (crosses, purple), $\kappa\to\infty$ (triangles, black). Thin solid lines with respective colours serve as a guide to the eye. Dashed lines with the same colour coding represent linear upper bounds in the form $N{\cal X}^{(1)}(L,\kappa)$.}
\end{figure*}

Using the numerical approach presented in the preceding section, we calculated contributions to Holevo quantity for sequences of $L=4$ and $L=5$ bins containing up to $N=7$ and $N=5$ photons, respectively, assuming binary encoding, $M=2$. The results are shown in figure \ref{fig:chiN}. For all numerical examples we found that the contribution from the $N$-photon sector is bounded from above by
\begin{equation}
\label{Eq:XN<=NX1}
{\cal X}^{(N)}(L, \kappa) \le N {\cal X}^{(1)} (L, \kappa).
\end{equation}
Additionally, we have also verified numerically this inequality for longer sequences, up to $L=22$ in the two-photon sector and up to $L=9$ in the three-photon sector.

Assuming that the inequality (\ref{Eq:XN<=NX1}) holds in a general case would allow us to obtain the following bound on the Holevo quantity:
\begin{equation}
{\cal X}(L,\kappa) \le \sum_{N=0}^{\infty} N P_N(L\bar{n}) {\cal X}^{(1)}(L,\kappa)  = L\bar{n} {\cal X}^{(1)} (L, \kappa).
\end{equation}
This in turn would give a universal bound on the accessible information per channel use in the form
\begin{equation}
{\cal I} = \lim_{L\rightarrow \infty} \frac{1}{L} {\cal X}(L,\kappa) \le \lim_{L\rightarrow \infty} \bar{n} {\cal X}^{(1)}(L, \kappa) = \bar{n} {\cal X}^{(1)}_{L\rightarrow\infty}(\kappa)
\end{equation}
valid for an arbitrary photon number $\bar{n}$. It is seen that the asymptotic value of the Holevo quantity in the one-photon sector plays the role of the proportionality factor between the accessible information and the average photon number per time bin. This factor, depicted in figure \ref{fig:chi1}, specifies the enhancement that can be gained from partial coherence between time bins that remains after dephasing. Its divergence in the limit $\kappa \rightarrow 0$ indicates the qualitative change of scaling from the linear one, as discussed in the introduction.

\begin{figure*}
\centering
\subfigure[]{
\label{fig:total1}\includegraphics[angle=0,width=0.47\textwidth]{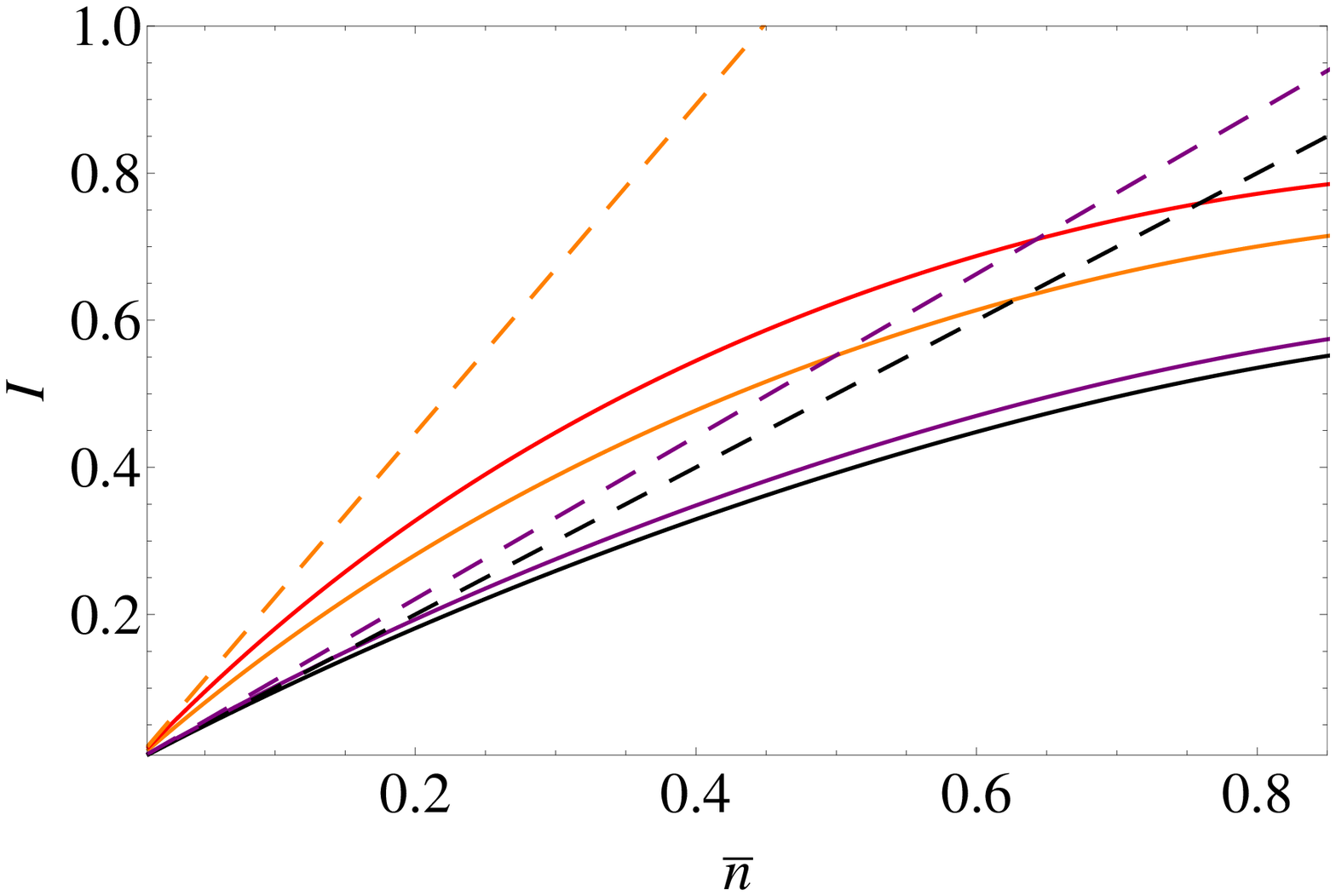}}
\subfigure[]{
\label{fig:total2}\includegraphics[angle=0,width=0.485\textwidth]{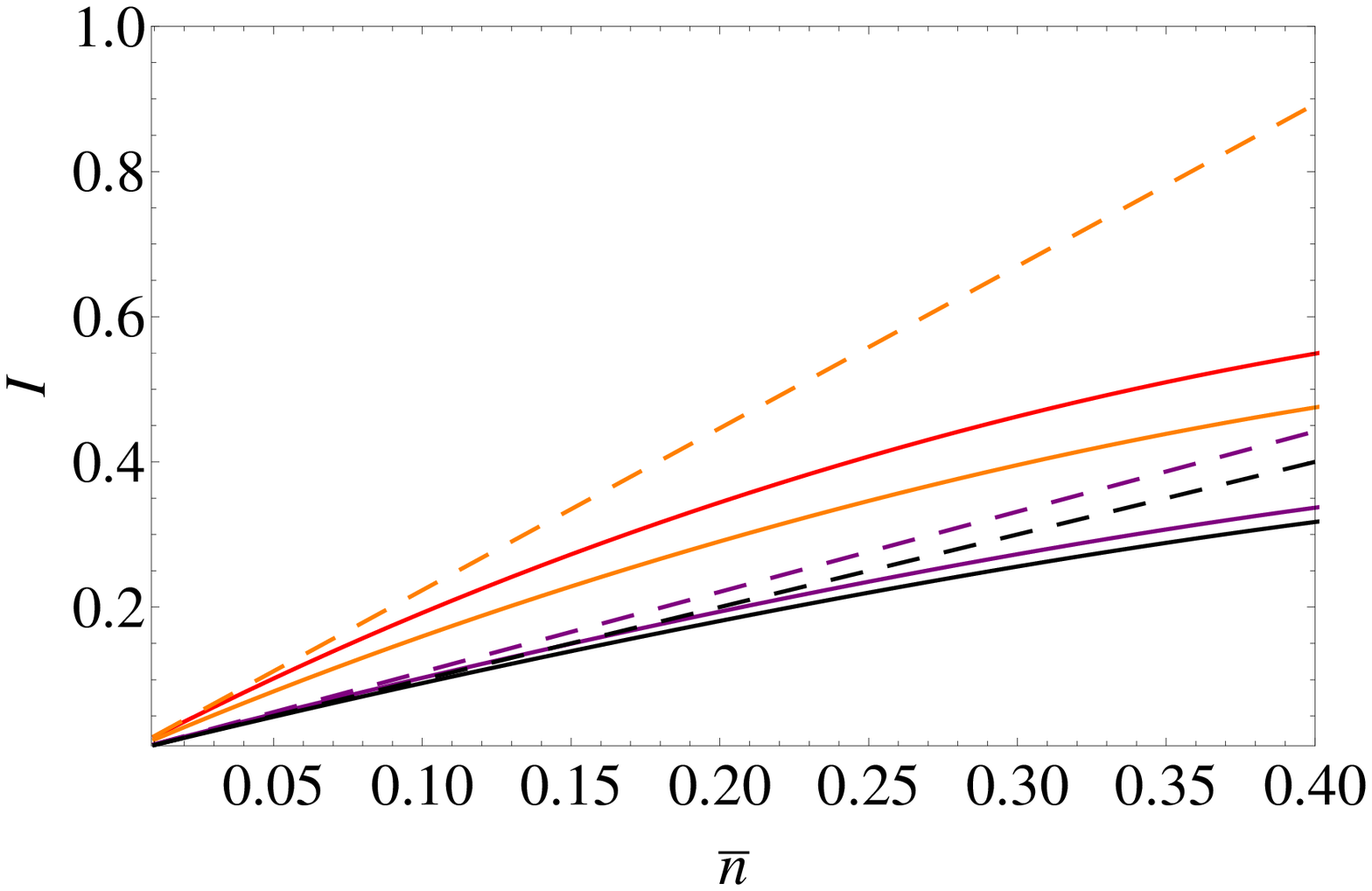}}
\caption{\label{fig:total}
Finite-length accesible information ${\cal X}(L,\kappa)/L$ (solid lines) and the conjectured asymptotic bounds $\bar{n}{\cal X}^{(1)}_{L\to\infty}(\kappa)$ (dashed lines) as a function of the average photon number $\bar{n}$ in a pulse for (a) $L=4$ and (b) $L=5$ time bins assuming binary encoding $M=2$ and the dephasing parameter $\kappa=0$ (red), $\kappa=0.1$ (orange), $\kappa=1$ (purple), and $\kappa\to\infty$ (black). Note that there is no linear bound
represented by a dashed line for $\kappa=0$ as $ {\cal X}^{(1)}_{L\to\infty}(0) = \infty$.}
\end{figure*}

In figure \ref{fig:total} we plot the accessible information as a function of the average photon number $\bar{n}$ calculated using the numerical values of contributions ${\cal X}^{(N)}(L,\kappa)$. In the graphs, we chose the range of $\bar{n}$ such that the Poisson distribution $P_N(L\bar{n})$ truncated at the highest photon number $N$ included in the calculations yields at least $95\%$ of the actual mean value $L\bar{n}$. The results are compared with the conjectured linear bound $\bar{n} {\cal X}^{(1)}_{L\rightarrow\infty}(\kappa)$. While for strong dephasing the linear bound matches the numerical results in the limit $\bar{n} \rightarrow 0$, for weak dephasing the bound becomes less tight. The principal reason for this change is that the coherence time of the channel is longer than the bin sequence included in numerical calculations.

%

\section{Photon splitting}
\label{Sec:CollectiveNoise}

We will close the paper by discussing one physically motivated attempt to prove inequality (\ref{Eq:XN<=NX1})  that illustrates
certain mathematical subtleties of the communication scheme considered here.
The basic idea is to apply a photon number splitting procedure analogous to eavesdropping in quantum key distribution with faint laser pulses \cite{lutkenhaus2000} to split deterministically $N$ photons prepared in the state $\ket{\Psi_{\bf m}^{(N)}}$ into $N$ separate paths, each containing exactly one photon. This realizes a transformation
\begin{equation}
\ket{\Psi_{\bf m}^{(N)}} \mapsto \ket{\Psi_{\bf m}^{(1)}} \otimes \ldots \otimes \ket{\Psi_{\bf m}^{(1)}},
\end{equation}
i.e.\ individual photons are described by the same state $\ket{\Psi_{\bf m}^{(1)}}$ and the overall state is given by its $N$-fold tensor product.
The photon splitting procedure can be in principle reversed using a suitably chosen unitary transformation \cite{lutkenhaus2000}, therefore communication using the ensemble of states $\hat{\varrho}_{\bf m}^{(N)} = \proj{\Psi_{\bf m}^{(N)}}$ should be equivalent to that using states $(\hat{\varrho}_{\bf m}^{(1)})^{\otimes N}$, where $\hat{\varrho}_{\bf m}^{(1)} = \proj{\Psi_{\bf m}^{(1)}}$.
In the absence of dephasing, the entropy of the average state can be estimated as
\begin{equation}
{\sf S}\left(  \sum_{\bf m} \frac{1}{M^L} (\hat{\varrho}_{\bf m}^{(1)})^{\otimes N} \right)
\le N {\sf S}\left( \sum_{\bf m} \frac{1}{M^L}  \hat{\varrho}_{\bf m}^{(1)}  \right)
\end{equation}
which follows immediately from the subadditivity of von Neumann entropy, while for an individual state we have
\begin{equation}
{\sf S}\bigl( \bigl( \hat{\varrho}_{\bf m}^{(1)} \bigr)^{\otimes N} \bigr) = N {\sf S}\bigl(  \hat{\varrho}_{\bf m}^{(1)} \bigr)
\end{equation}
as individual states are completely uncorrelated. These two identities combined together give the sought bound
\begin{eqnarray}
{\cal X}^{(N)} =
{\sf S}\left(  \sum_{\bf m} \frac{1}{M^L} (\hat{\varrho}_{\bf m}^{(1)})^{\otimes N} \right)
 - \sum_{\bf m} \frac{1}{M^L}
{\sf S}\bigl( \bigl( \hat{\varrho}_{\bf m}^{(1)} \bigr)^{\otimes N} \bigr) \nonumber \\
 \le N \left[  {\sf S}\left( \sum_{\bf m} \frac{1}{M^L}  \hat{\varrho}_{\bf m}^{(1)}  \right)
-  \sum_{\bf m} \frac{1}{M^L} {\sf S}\bigl(  \hat{\varrho}_{\bf m}^{(1)} \bigr)
\right] = N {\cal X}^{(1)}
\label{Eq:IneqIndependentDephasing}
\end{eqnarray}
for the fully coherent case.
However, the above argument no longer works in the presence of dephasing. The reason is that dephasing applied to the state $\ket{\Psi_{\bf m}^{(N)}}$ introduces phase correlations between photons after splitting them into separate paths and their joint state cannot be represented as a tensor product of identical individual states. To illustrate this point, let us consider the case of $N=1$ and $N=2$ photons in two time bins. Let the word ${\bf m}$ be fixed, which allows us to consider only one field mode per time bin. The single-photon sector state after dephasing written in the Fock basis $\{ \ket{0,1}, \ket{1,0}\}$ takes the form
\begin{equation}
\label{Eq:onephotondephased}
\Lambda (\proj{\Psi_{\bf m}^{(1)}}) = \frac{1}{2}
\left( \begin{array}{cc} 1 & e^{-\kappa} \\
e^{-\kappa} & 1 \end{array} \right).
\end{equation}
Two independently dephased single photons are described in the product Fock state basis by the joint state:
\begin{equation}
\label{Eq:onetimesone}
\Lambda (\proj{\Psi_{\bf m}^{(1)}}) \otimes \Lambda (\proj{\Psi_{\bf m}^{(1)}})
=
\frac{1}{4} \left( \begin{array}{cccc}
1 & e^{-\kappa} & e^{-\kappa} & e^{-2\kappa} \\
e^{-\kappa} & 1 & e^{-2\kappa} & e^{-\kappa} \\
e^{-\kappa} & e^{-2\kappa} & 1 & e^{-\kappa} \\
e^{-2\kappa} & e^{-\kappa} & e^{-\kappa} & 1
\end{array} \right).
\end{equation}
On the other hand, consider the input state in the two-photon sector
\begin{equation}
\ket{\Psi_{\bf m}^{(2)}} = \frac{1}{\sqrt{2}} \left(\frac{\hat{a}_{1m_1}^\dagger + \hat{a}_{2m_2}^\dagger}{\sqrt{2}} \right)^2 \ket{\text{vac}}
= \frac{1}{2} \ket{2,0} + \frac{1}{\sqrt{2}} \ket{1,1} + \frac{1}{2} \ket{0,2}.
\end{equation}
When the two photons are subjected to dephasing and split into separate paths according to a map
\begin{eqnarray}
\ket{2,0} \mapsto \ket{1,0} \otimes \ket{1,0}  \nonumber \\
\ket{1,1} \mapsto \frac{1}{\sqrt{2}} (\ket{1,0} \otimes \ket{0,1} + \ket{0,1} \otimes \ket{1,0}) \\
\ket{0,2} \mapsto \ket{0,1} \otimes \ket{0,1}  \nonumber
\end{eqnarray}
their final state written in the same basis as in (\ref{Eq:onetimesone}) takes the form
\begin{equation}\label{eq:two}
\Lambda(\proj{\Psi_{\bf m}^{(2)}}) \mapsto
\frac{1}{4} \left( \begin{array}{cccc}
1 & e^{-\kappa} & e^{-\kappa} & e^{-4\kappa} \\
e^{-\kappa} & 1 & 1 & e^{-\kappa} \\
e^{-\kappa} & 1 & 1 & e^{-\kappa} \\
e^{-4\kappa} & e^{-\kappa} & e^{-\kappa} & 1
\end{array} \right).
\end{equation}
This expression is clearly different from (\ref{Eq:onetimesone}), although tracing over one of the two photons does reproduce the dephased single photon state from (\ref{Eq:onephotondephased}). While the inequality (\ref{Eq:IneqIndependentDephasing}) holds also for states $\hat{\varrho}_{\bf m}^{(1)} = \Lambda(\proj{\Psi_{\bf m}^{(1)}})$, the above example shows that unfortunately in the general case the dephased multiphoton states $\Lambda(\proj{\Psi_{\bf m}^{(N)}})$ are not mapped by the beam splitting procedure onto $N$-fold tensor products
$[\Lambda(\proj{\Psi_{\bf m}^{(N)}})]^{\otimes N}$.

\begin{figure*}
\centering
\subfigure[]{
\label{fig:holentr1}\includegraphics[angle=0,width=0.485\textwidth]{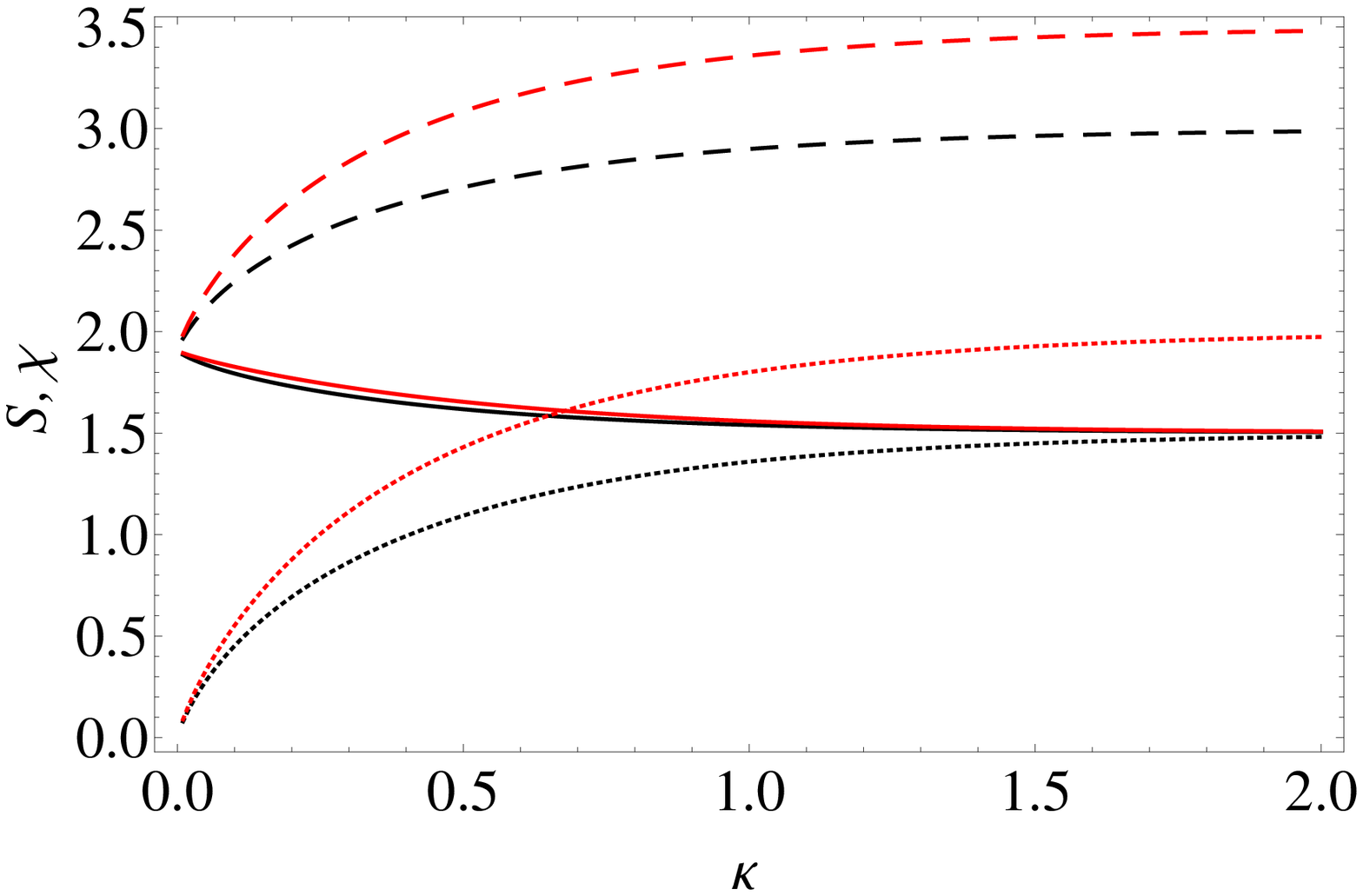}}
\subfigure[]{
\label{fig:holentr2}\includegraphics[angle=0,width=0.47\textwidth]{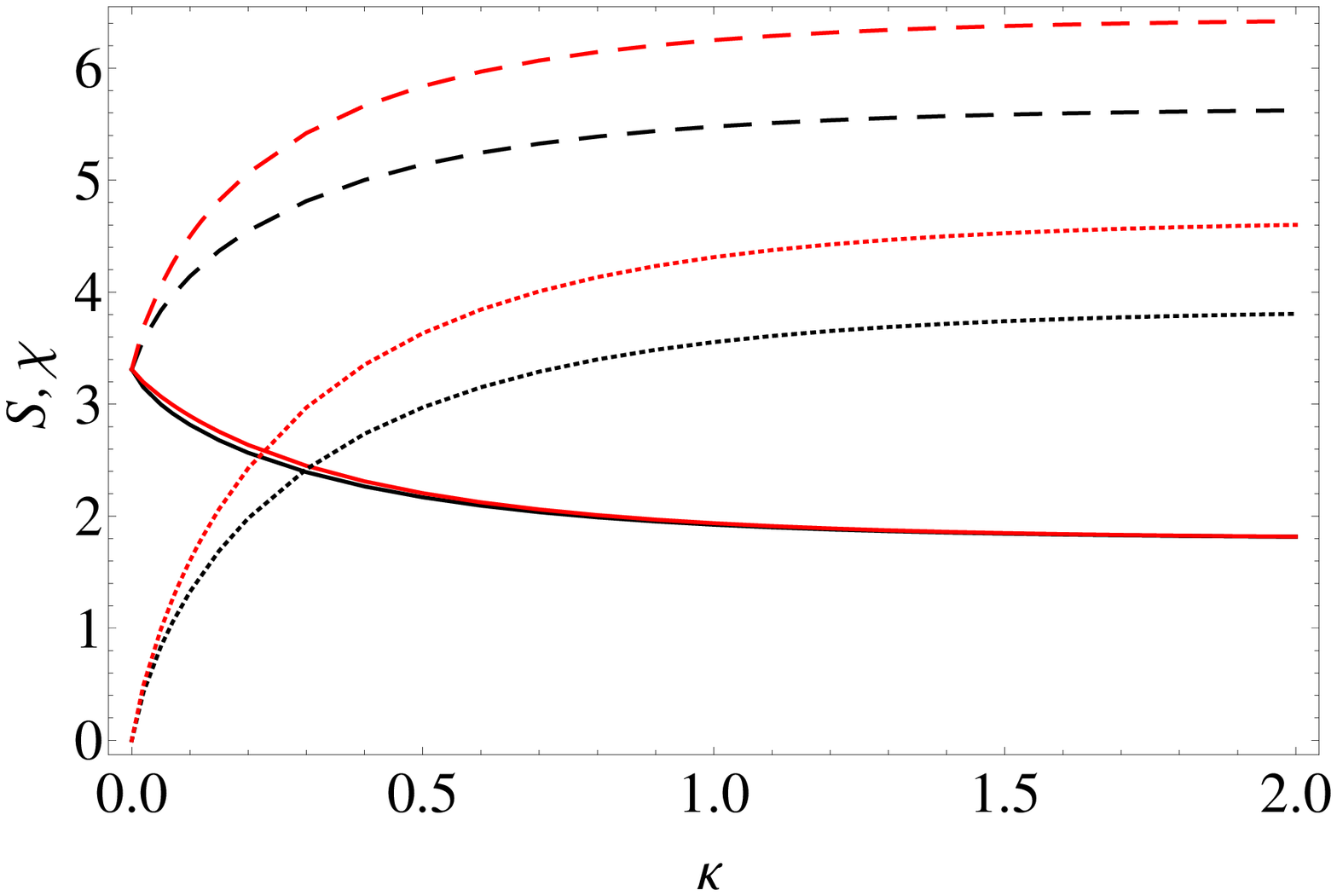}}
\caption{\label{fig:holentr}
The two-photon contribution to the Holevo quantity (solid line), the entropy of the average state in the two-photon sector (dashed line), and the entropy of an individual state in the two-photon sector (dotted line) as a function of the parameter $\kappa$ for a pair of photons assuming independent dephasing (red colour) and collective dephasing (black colour) for binary encoding $M=2$ and (a) $L=2$ and (b) $L=5$ time bins.}
\end{figure*}

The difference between independent and collective dephasing for photons split into separate paths has consequences for the Holevo quantity. In figure \ref{fig:holentr} we plot as a function of $\kappa$  entropies of individual and average states in the two-photon sector as well as contributions to the Holevo quantities ${\cal X}^{(2)}$ given by the difference between these two, assuming either the actual photon pairs subjected to collective dephasing or two independently dephased single photons. The case of binary coding $M=2$ and sequences of either $L=2$ or $L=5$ time bins is shown in the graphs. In these plots, the Holevo quantity for collective dephasing turns out to be slightly smaller than its counterpart for individual dephasing. If this observation was a universal feature, it would pave the way to prove the inequality conjectured in (\ref{Eq:XN<=NX1}) in the general case using (\ref{Eq:IneqIndependentDephasing}).
However, its validity seems to depend heavily on the specifics of the input states and the noise model considered in this work rather than to follow from general properties of noisy quantum channels.

\section{Conclusions}
\label{Sec:Conclusions}

We have analysed the accessible information of an optical communication scheme based on sending a sequence of symbols encoded as coherent states of orthogonal field modes. We implemented a numerical procedure to compute the Holevo quantity and analysed analytically the weak pulse limit, when the number of photons in a time bin sequence is much less than one. The obtained results suggest a general fact that if consecutive transmissions experience dephasing, the Holevo quantity scales linearly with the average photon number in the limit of very weak pulses. This is a qualitative departure from the fully coherent scheme, where the Holevo quantity becomes enhanced by a logarithmic factor. Nevertheless, even partial preservation of the relative phase can improve the communication capacity compared to incoherent direct detection by increasing the multiplicative factor that relates the average photon number to the accessible information. This conclusion resembles the results obtained recently in quantum metrology \cite{rdd}, where in many scenarios even arbitrarily weak decoherence changes dependence of the measurement precision on the number of probes used from the Heisenberg scaling to the shot-noise scaling, albeit the latter can be improved by a multiplicative factor if the probes are prepared and measured collectively.

The current work leads to questions about the practical attainability of the Holevo limit for communication over realistic channels with phase noise. This includes design of receivers, efficient schemes for calibrating the phase reference, and the sensitivity of optimal strategies to the specific form of dephasing and other noise mechanisms occurring in realistic scenarios. In contrast to previous studies of classical communication over correlated noise channels \cite{corr1,corr2,corr3} optimization over input states is not attempted here, limiting attention to coherent states that can be prepared using standard techniques. However, even in such restricted settings, construction of receivers approaching the quantum limit remains a subject of ongoing research
\cite{receiver1,receiver2}.

\section*{Acknowledgements}

This research was supported by the Foundation for Polish Science TEAM project cofinanced by the EU European Regional Development Fund and the ERA-NET CHIST-ERA project QUASAR.


\appendix

\section{}
\label{Sec:AppA}
In order to evaluate ${\cal X}^{(1)}(L,\kappa)$ given in (\ref{Eq:X1sumh(t_i)}) in the asymptotic $L\rightarrow \infty$ we need to calculate the integral
\begin{equation}
I = \frac{1}{2\pi} \int_{-\pi}^{\pi} d\theta \, f(\theta) \log_2 f(\theta)
\end{equation}
with $f(\theta)$ defined in (\ref{Eq:f(theta)}).
It will be convenient to decompose $I = I_1 - (1-e^{-2\kappa}) I_2 / 2 \pi \ln 2$, where the integral
\begin{equation}
I_1 = \frac{1-e^{-2\kappa}}{2\pi}\int_{-\pi}^{\pi}d\theta\frac{\log_2(1-e^{-2\kappa})}{1+e^{-2\kappa}-2e^{-\kappa}\cos\theta}
= \log_2(1-e^{-2\kappa})
\end{equation}
can be evaluated by elementary means, while the integral $I_2$ can be transformed by substituting $e^{-\kappa}=t$ to the form
\begin{equation}
I_2= \int_{-\pi}^{\pi} d\theta
\frac{\ln (1+ e^{-2\kappa} - 2 e^{-\kappa} \cos\theta)}{1+ e^{-2\kappa} - 2 e^{-\kappa} \cos\theta} =
\int_{-\pi}^{\pi}d\theta\frac{\ln(1+t^2-2t\cos\theta)}{1+t^2-2t\cos\theta}
\end{equation}
We will now use expansion into Chebyshev polynomials of the first kind $T_n(x)$ and the second kind $U_n(x)$:
\begin{eqnarray}
\frac{1}{1+t^2-2t\cos\theta} = \sum_{n=0}^{\infty}U_n(\cos\theta)t^n\\
\frac{1}{2}\ln(1+t^2-2t\cos\theta) = - \sum_{n=1}^{\infty}T_n(\cos\theta)\frac{t^n}{n}
\end{eqnarray}
Inserting these expansions into the integral over $\theta$ and making use of the properties
of Chebyshev polynomials \cite{abramowitz} gives:
\begin{eqnarray}
\fl
\nonumber
I_2=-2\int_{-\pi}^{\pi}d\theta\sum_{n=1}^{\infty}\sum_{m=0}^{\infty}T_n(\cos\theta)U_m(\cos\theta)\frac{t^{n+m}}{n}\\
\nonumber
=-\int_{-\pi}^{\pi}d\theta\sum_{m=1}^{\infty}\sum_{n=1}^{m}\left(U_{n+m}(\cos\theta)+U_{m-n}(\cos\theta)\right)\frac{t^{n+m}}{n}\\
=-\sum_{m=1}^{\infty}\sum_{n=1}^{m}\int_{-\pi}^{\pi}d\theta\left(\frac{\sin[(n+m+1)\theta]}{\sin\theta}+\frac{\sin[(m-n+1)\theta]}{\sin\theta}\right)\frac{t^{n+m}}{n}
\end{eqnarray}
But $\int_{-\pi}^{\pi}d\theta \frac{\sin n\theta}{\sin\theta}=2\pi$ when $n$ is odd, otherwise the integral is equal to $0$. By changing summation variables to $2k+1=n+m+1$ and  $2l+1=m-n+1$ we obtain
\begin{eqnarray}
I_2=-4\pi\sum_{k=1}^{\infty}\sum_{l=0}^{k-1}\frac{t^{2k}}{k-l}=-4\pi\sum_{k=1}^{\infty}H_{k}t^{2k}=4\pi\frac{\ln(1-t^2)}{1-t^2}.
\end{eqnarray}
where $H_{k}$ is the $k$th harmonic number defined as $H_k=\sum_{n=1}^{k}\frac{1}{n}$. This result gives us an explicit formula for the sought integral:
\begin{equation}
I = \frac{1}{2\pi} \int_{-\pi}^{\pi} d\theta \, f(\theta) \log_2 f(\theta)=-\log_2(1-e^{-2\kappa}).
\end{equation}


\providecommand{\newblock}{}

\end{document}